\documentclass[%
 reprint,amsmath,amssymb,aps
]{revtex4-2}

\usepackage{fancyhdr}

\usepackage{graphicx}
\usepackage[dvipsnames]{xcolor}
\definecolor{darkblue}{RGB}{25,88,128}
\usepackage[
    colorlinks=true,
    linkcolor=darkblue,
    filecolor=darkblue,
    citecolor=darkblue,      
    urlcolor=darkblue,
    ]{hyperref}

\begin{document}

\title{Influence of germanium substitution on the structural and electronic stability of the competing vanadium dioxide phases}

\author{Peter Mlkvik}\email{peter.mlkvik@mat.ethz.ch}
\author{Claude Ederer}
\author{Nicola A. Spaldin}
\affiliation{Materials Theory, Department of Materials, ETH Z\"{u}rich, Wolfgang-Pauli-Strasse 27, 8093 Z\"{u}rich, Switzerland}

\date{\today}

\begin{abstract}
We present a density-functional theory (DFT) study of the structural, electronic, and chemical bonding behavior in germanium (Ge)-doped vanadium dioxide (VO$_2$). Our motivation is to explain the reported increase of the metal-insulator transition temperature under Ge doping and to understand how much of the fundamental physics and chemistry behind it can be captured at the conventional DFT level. We model doping using a supercell approach, with various concentrations and different spatial distributions of Ge atoms in VO$_2$. Our results suggest that the addition of Ge atoms strongly perturbs the high-symmetry metallic rutile phase and induces structural distortions that partially resemble the dimerization of the experimental insulating structure. Our work, therefore, hints at a possible explanation of the observed increase in transition temperature under Ge doping, motivating further studies into understanding the interplay of structural and electronic transitions in VO$_2$.
\end{abstract}

\maketitle
\thispagestyle{fancy}
\section{Introduction}
\begin{figure}
    \centering
    \includegraphics[width=1\linewidth]{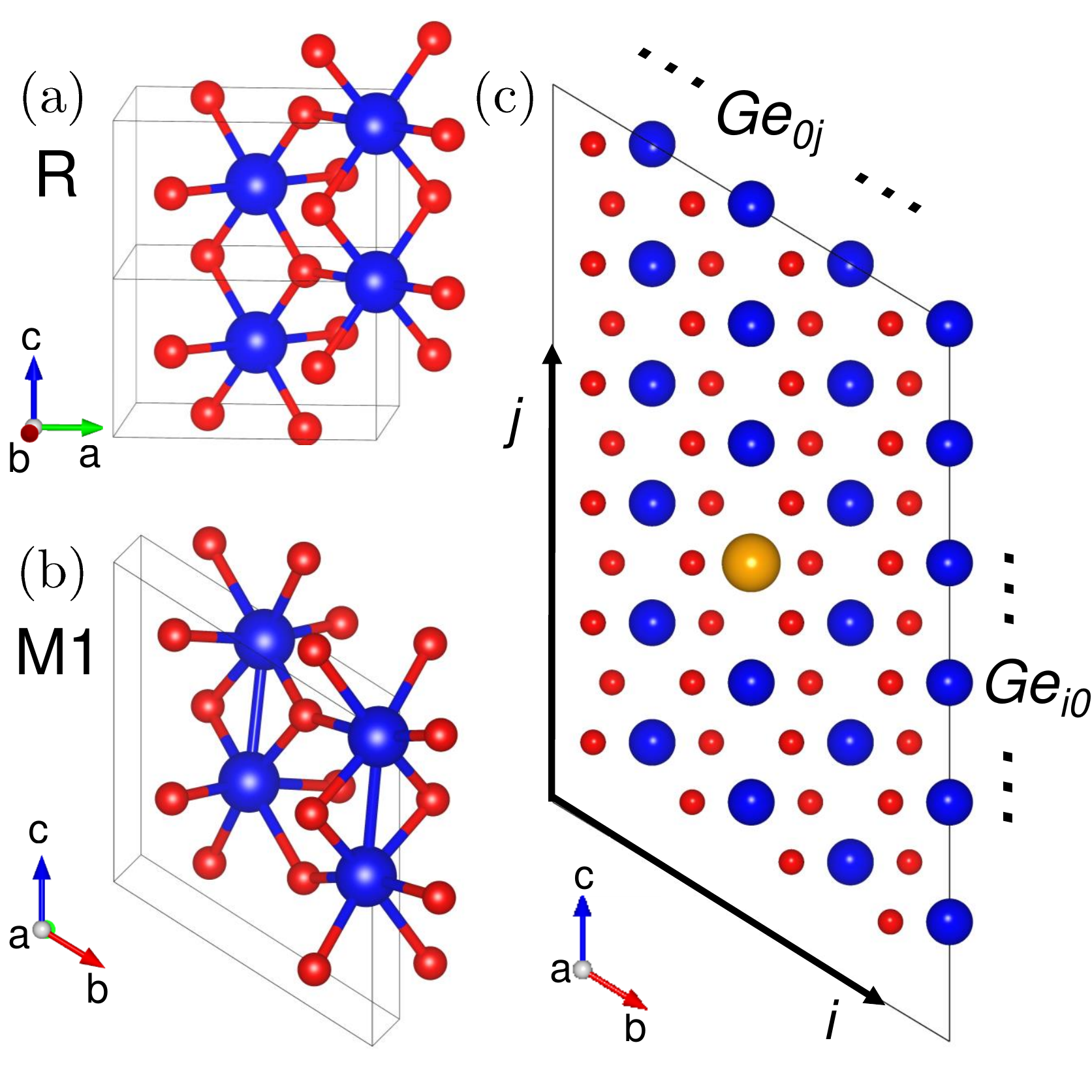}
    \caption{(a)~Double unit cell of the R and (b)~single unit cell of M1 VO$_2$ phases in their pristine condition. (c)~R phase $2 \times 2 \times 3$ supercell used for calculations with the stationary Ge atom (orange) corresponding to the Ge$_{00}$ coordinate of the labeling system indicated. V (O) atoms shown in blue (red).}
    \label{fig:structures}
\end{figure}

Vanadium dioxide (VO$_2$) is a prototypical example of a system undergoing a metal-insulator transition (MIT) coupled with a structural transition. This first-order transition from a high-temperature rutile (R) to a low-temperature monoclinic (M1) phase (Figs.~\ref{fig:structures} (a) and (b)) spans several orders of magnitude in resistivity~\cite{Eyert:2002}, making VO$_2$ an interesting target for many potential uses~\cite{Yang/Ko/Ramanathan:2011, Ke_et_al:2018, Liu_et_al:2018, Andrews_et_al:2019, Corti_et_al:2021}. Although the transition temperature of around $T_c$~=~340~K is close to room temperature, its tuning has been a major focus of research~\cite{Shao_et_al:2018, Shi_et_al:2019}. Its increase could lead to possibilities in the sector of electrical switches~\cite{Rini_et_al:2008, Vitale_et_al:2016, Shao_et_al:2018, Rosca_et_al:2021}, and its decrease could lead to applications in smart windows~\cite{Wang_et_al:2016, Cui_et_al:2018}. In this study, we focus on understanding the behavior under doping with germanium (Ge), which has recently been reported to increase the transition temperature~\cite{Krammer_et_al:2017, Muller_et_al:2020}. Specifically, we perform calculations using density-functional theory (DFT) to determine the structural, electronic, and chemical effects that Ge doping has on VO$_2$ at the DFT level.
 
The MIT in VO$_2$ has been widely studied in the last decades with discussions mainly about whether structural or electronic effects provide the leading physical mechanism driving the transition~\cite{Zylbersztejn/Mott:1975,  Wentzcovitch/Schulz/Allen:1994, Biermann_et_al:2005, Eyert:2011, Weber_et_al:2012, Grau-Crespo/Wang/Schwingenschlogl:2012, Kim_et_al:2013, Brito_et_al:2016, Najera_et_al:2017, Grandi/Amaricci/Fabrizio:2020, Pouget:2021}. Structurally, VO$_2$ undergoes a dimerization  of V atoms along the crystal $c$ axis (Figs. \ref{fig:structures}, (a) and (b)), coupled with tilting of these dimers away from $c$ upon lowering the temperature through $T_c$. This mechanism has initially been attributed to Peierls physics leading in turn to a band-gap opening~\cite{Goodenough:1971}. Thereby, the $a_{1g}$ orbital (also called $d_{||}$, due to its lobe being oriented along the chain direction, see Ref.~\cite{Eyert:2002}), occupied with the single $d$ electron of the V$^{4+}$ cation, splits during the transition into a bonding-antibonding pair. However, the transition is unlikely to originate only from such structural effects, and calculations using conventional DFT indicate that purely structural deformations do not cause the formation of a gap~\cite{Eyert:2002}. Additionally, an insulating M2 phase containing both dimerized chains and non-dimerized tilted chains exists under certain conditions~\cite{Quackenbush_et_al:2016, Davenport_et_al:2021}, showing that a gap is allowed to form even when half of the V atoms are equally spaced~\cite{Pouget_et_al:1974}. Hence, the correlation effects among the localized $d$ electrons seem to also play an important role in the material, motivating numerous dynamical mean-field theory (DMFT)~\cite{Tomczak/Biermann:2007, Tomczak/Aryasetiawan/Biermann:2008, Belozerov_et_al:2012, Brito_et_al:2016, Brito_et_al:2017} and $GW$-based studies~\cite{Gatti_et_al:2007, Weber_et_al:2020}. 

In this work, we study the behavior of VO$_2$ when some V atoms are replaced with Ge atoms. Although charged dopants in VO$_2$ have been experimentally~\cite{Asayesh-Ardakani_et_al:2015, Davenport_et_al:2021} and computationally~\cite{Stahl/Bredow:2021} studied before, and seem to follow a trend of increased $T_c$ with increased valence of the dopant~\cite{Shi_et_al:2019}, the role of charge-neutral dopants such as Ge remains unclear, with few computational studies aiming at explaining the effects under doping using DFT~\cite{Chen_et_al:2017, Lu/Guo/Robertson:2019, Lu_et_al:2020, Lu_et_al:2021}. Chen {\it et al.}~\cite{Chen_et_al:2017} conducted a large-scale DFT study of the M1 structure of VO$_2$ doped with group IV elements including Ge, focusing on absorption and reflectivity, predicting a decrease in $T_c$ for all tested elements, at odds with experiments. Lu {\it et al.}~\cite{Lu/Guo/Robertson:2019} conducted a DFT study of VO$_2$ alloying with 25\% GeO$_2$, comparing it with MgO$_2$ alloying. The authors observed that the heavily Ge-doped VO$_2$ collapses from the M1 phase to the R phase due to GeO$_2$ being rutile, and they focused on magnetic ordering in the material. The same authors~\cite{Lu_et_al:2020} later considered lower Ge doping concentrations, focusing on the dependence of magnetic ordering and enthalpy changes on various density-functional methodologies. Using non-collinear supercell-based paramagnetic calculations for the R phase combined with a Heisenberg dimer model, they found an increase in $T_c$ under Ge doping in agreement with experiments.

Here, we build on these earlier works to present a combined structural, electronic, and chemical picture of the effects a Ge dopant has on VO$_2$. We focus particularly on the effect of Ge doping on the structure of VO$_2$, as well as on the electronic behavior of V atoms both close to and far away from the dopants, considering multiple different configurations of the dopant atoms. In addition, we consider the chemical bonding behavior of the Ge-V and V-V atom pairs near the dopant. We evaluate the structural response of the two main VO$_2$ phases (R and M1) to Ge doping and discuss whether this response can help explain experimental observations. The second key goal of our work is to assess to what extent the standard DFT methodology employed here and elsewhere~\cite{Eyert:2002} is sufficient to model this behavior.

We expect the effects of Ge dopant atoms to come from two main sources.
Firstly, Ge and V have different atomic radii and oxygen octahedra sizes, so substituting V with Ge will lead to structural changes from steric effects. Secondly, Ge and V have different numbers of valence electrons, so the bonding behavior of a Ge atom could be significantly different from that of a V atom. We hence look at the structural, electronic, and chemical bonding in the doped systems.

\fancyhf{}
\renewcommand{\headrulewidth}{0pt}
\rhead{\thepage}
\section{Methods}

\subsection{Construction of the unit cells}
\label{sec:supercell}

Structural relaxations are performed for both the low-temperature M1 phase and the high-temperature R phases with monoclinic and tetragonal primitive unit cells respectively. In their primitive cells (Figs.~\ref{fig:structures} (a) and (b)), the R structure contains two VO$_2$ formula units, with the shortest distance between neighboring V atoms along the $c$~axis (a doubled unit cell of R is shown in Fig.~\ref{fig:structures} (a) for easier comparison with the M1 phase in Fig.~\ref{fig:structures} (b)). The M1 structure contains four VO$_2$ formula units and is formed by doubling the R cell along its $c$~axis (and a monoclinic shear leading to a redefined $b$ axis). In the current work, we align the M1 $c$ axis parallel to the one of the R structure. For doped calculations, we consider two supercells; $2 \times 2 \times 2$ and $2 \times 2 \times 3$ relative to the M1 unit cell (the latter tripled in the chain direction, Fig.~\ref{fig:structures} (c)), containing 32 and 48 formula units respectively.

The typical doping range for Ge-doped VO$_2$ is between 2\% and 6\%~\cite{Krammer_et_al:2017, Muller_et_al:2021}, hence we consider either one or two Ge atoms per supercell.
For doping with a single Ge atom, only one inequivalent site exists in both the $2 \times 2 \times 2$ and $2 \times 2 \times 3$ supercells, corresponding to a dopant concentration of 3.125\% and 2.08\%, respectively. For the $2 \times 2 \times 3$ case we also consider doping with two Ge atoms, corresponding to a dopant level of 4.17\%.
In this case, there are many distinct sets of possible positions of the two Ge dopants in the R and M1 structures.

In order to capture the disorder present in doped VO$_2$, we probe different possible configurations of the two Ge atoms in the $2 \times 2 \times 3$ supercell. Thereby, the Ge positions are labeled according to their projections along the $\vec{a}$~lattice vector, as shown in Fig.~\ref{fig:structures} (c), as Ge$_{ij}$. The labeling system follows the axes of the supercell, i.e. the M1 lattice vectors $\vec{b}$ and $\vec{c}$. The first index  $i$, labels successive chains, with $i$ increasing along the positive $\vec{b}$ lattice vector, and the second index $j$ labels the positions along the chains, increasing along the positive $\vec{c}$ lattice vector. All investigated configurations have the first Ge atom located at the origin of the labeling system (i.e. Ge$_{00}$, orange circle in Fig.~\ref{fig:structures}(c)) and are labeled according to the position of the second Ge atom. Thus, a configuration is labeled as Ge$_{0j}$ if the second Ge atom is located within the same chain, as Ge$_{1j}$ if it is located in a nearest-neighboring chain, and as Ge$_{2j}$ if it is in a second nearest-neighboring chain (with $j=-2,\,...\,3$). Since we only consider configurations including Ge atoms in one first nearest-neighbor chain and one second nearest-neighbor chain, we do not need to report the position along $\vec{a}$. Note that the second nearest-neighbor chains are at zero $\vec{a}$, while the nearest-neighbor chains are halfway along the unit cell in the $\vec{a}$ direction. For the R structure, considering up to the second nearest-neighbor chain, this combination yields 10 possible configurations all of which we calculate. For the M1 structure, there are 53 possible configurations, but due to computational considerations, we only treat a representative sample of 17 of these, in particular, one set of those of the R structure, adding the configurations along the chains, Ge$_{i-1}$, Ge$_{i-2}$, and Ge$_{i-3}$.

\subsection{Computational Details}

We perform calculations using DFT with plane-wave basis sets as implemented in the \textsc{quantum espresso} code (QE v6.4.1)~\cite{Giannozzi_et_al:2009, Giannozzi_et_al:2017}. Calculations are performed within the generalized gradient approximation (GGA) using the Perdew-Burke-Ernzerhof (PBE)~\cite{Perdew/Burke/Ernzerhof:1996} exchange-correlation functional. Note that since our goal is to capture the effects of the doping chemistry, and we do not aim to reproduce the MIT or Mott physics, we perform non-spin polarized calculations, without a DMFT or $GW$ treatment and also without a DFT+$U$ correction~\cite{Stahl/Bredow:2020}. Using our setup, the lattice and cell sizes, together with the qualitative behavior of the density of states (DOS) of the two phases, are captured correctly at a computational cost that is suitable for the supercell calculations. Note that, as in previous standard DFT calculations~\cite{Eyert:2011}, we do not obtain a band gap for the insulating M1 phase, and the two VO$_2$ phases have the incorrect relative energies~\cite{Mellan_et_al:2019} (each having its own stable minimum), but since we treat each phase separately, these features do not pose problems for our analysis. 

We use scalar-relativistic ultrasoft pseudopotentials for O, V, and Ge atoms taken from the SSSP library~\cite{Lejaeghere_et_al:2016, Prandini_et_al:2018}, with 3$s$ and 3$p$ semi core states included as valence in V atoms, and 3$d$ states included as valence for Ge atoms. Calculations are conducted with a wavefunction plane-wave kinetic energy cut-off of 40~Ry ($\sim$ 544~eV) increased by a factor of 8 for the charge density and a $4 \times 4 \times 4$ $\Gamma$-centered $k$-point grid for all calculations. We converge the total energies to 10$^{-6}$~eV and in structural relaxations we converge all force components to 10$^{-3}$~eV/Å. During structural relaxation, both the internal positions of the atoms and the lattice vectors are relaxed. We observe only very small ($\sim$0.01\%) lattice parameter changes relative to the pristine structures.

We perform a band unfolding procedure to obtain a primitive cell band structure from the supercell folded band structure. We follow the procedure outlined in Ref.~\cite{Popescu/Zunger:2012} (and previously in Refs.~\cite{Boykin/Klimeck:2005, Boykin_et_al:2007}), implemented in the \textsc{bandup} code~\cite{Medeiros/Stafstrom/Bjork:2014, Medeiros_et_al:2015} through the \textsc{banduppy}~\cite{Iraola_et_al:2022} \textsc{python} interface. To determine the chemical bonding behavior from the obtained DFT results, the crystal orbital hamiltonian population (COHP) analysis~\cite{Dronskowski/Bloechl:1993} is performed using the \textsc{lobster} package~\cite{Deringer/Tchougreeff/Dronskowski:2011, Maintz_et_al:2013, Maintz_et_al:2016}.


\section{Results and Discussion}

\begin{figure*}
    \centering
    \includegraphics[width=1\textwidth]{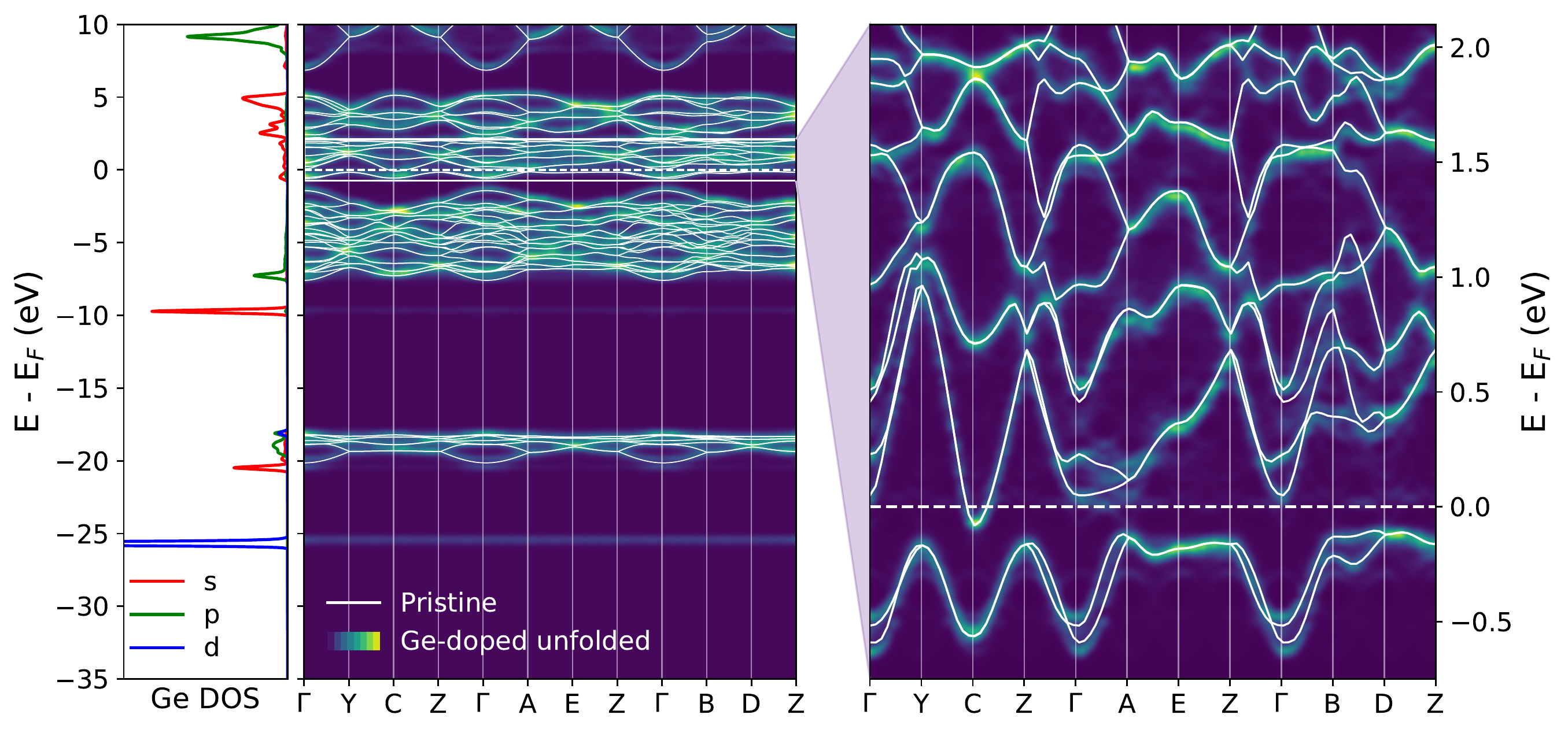}
    \caption{Band structure of the Ge-doped M1 phase with the corresponding local DOS of the Ge dopant. In white, the pristine band structure is overlaid over the unfolded Ge-doped band structure, shown through a color map. The certainty of the unfolded energy eigenvalues corresponds to the color intensity. The Fermi level (dashed white line) is set to the zero of energy. Left panel: The Ge DOS shows $s$ (red), $p$ (green), and $d$ (blue) orbitals. Center panel: Band structure of the Ge-doped M1 phase. Right panel: Detail of the band structure around the Fermi level showing the pristine bands and the unfolding dispersion due to Ge states.}
    \label{fig:bandstructure}
\end{figure*}

\subsection{VO$_2$ cells containing an isolated Ge atom}

\subsubsection{Electronics}

We first perform self-consistent calculations for the $2 \times 2 \times 2$ unit cell with a single Ge atom and structural parameters fixed to those obtained for pristine M1 VO$_2$. This allows us to analyze the changes in the electronic structure that are solely due to the different chemistry of the Ge atom, without the potential structural changes. 

The corresponding Ge DOS and unfolded band structure are compared to those of the undoped structure in Fig.~\ref{fig:bandstructure}. The latter agrees well with previous DFT studies~\cite{Eyert:2002, Sakuma/Miyake/Aryasetiawan:2009, Eyert:2011, Yuan_et_al:2012}. The O~$2s$~bands are located at around $-$20~eV and the O~$2p$-dominated bands (with some contribution of V~$d$~states due to hybridization) lie between $-$8 and $-$1.5~eV. The Fermi level cuts through the bottom of the V~$3d$-dominated bands (also containing some O~$2p$ character due to hybridization). Due to the octahedral coordination of the V cations, the V~$3d$ band is separated into lower-lying $t_{2g}$-like (from $-$0.5 to around 2.5~eV) and higher-lying $e_g$-like (from around 2.5 to 5~eV) states. 

From the projected DOS shown in the left-most panel of Fig.~\ref{fig:bandstructure}, one can see that the Ge states in the doped structure are mostly located far away from the Fermi level. In particular, the completely filled Ge~$3d$ shell lies at around $-$26~eV. The Ge~$4s$ peak at $-$10~eV corresponds to a strong hybridization with an O~$2p$ band (not shown), introducing no carriers to the system, i.e., the Ge atom remains formally Ge$^{4+}$.

Thanks to the unfolding procedure, we can directly see the effect of the addition of a Ge atom on the band structure (Fig.~\ref{fig:bandstructure}, center and right panels).
Here, the larger the spread of a given energy eigenvalue is, the more its value varies across the different unit cells of the supercell, and hence the more it has been altered by the added Ge compared to pristine VO$_2$. The effects of Ge doping are shown through the color map smearing, with a brighter yellow color corresponding to a greater weight coming from the unfolding procedure. Most of the band energies remain in the same position as in pristine VO$_2$, with only minor shifts and smearing. There is also no noticeable change in the filling of the V~$d$~states. We note that the same behavior persists even after full structural relaxation, after which the unfolded bands become more diffuse due to the variable positions of atoms in the supercell. We hence see that the substitution of V by Ge has very little effect on the electronic states of the system.

\subsubsection{Chemical bonding}
\label{sec:chemistry}

\begin{figure}
    \centering
    \includegraphics[width=1\linewidth]{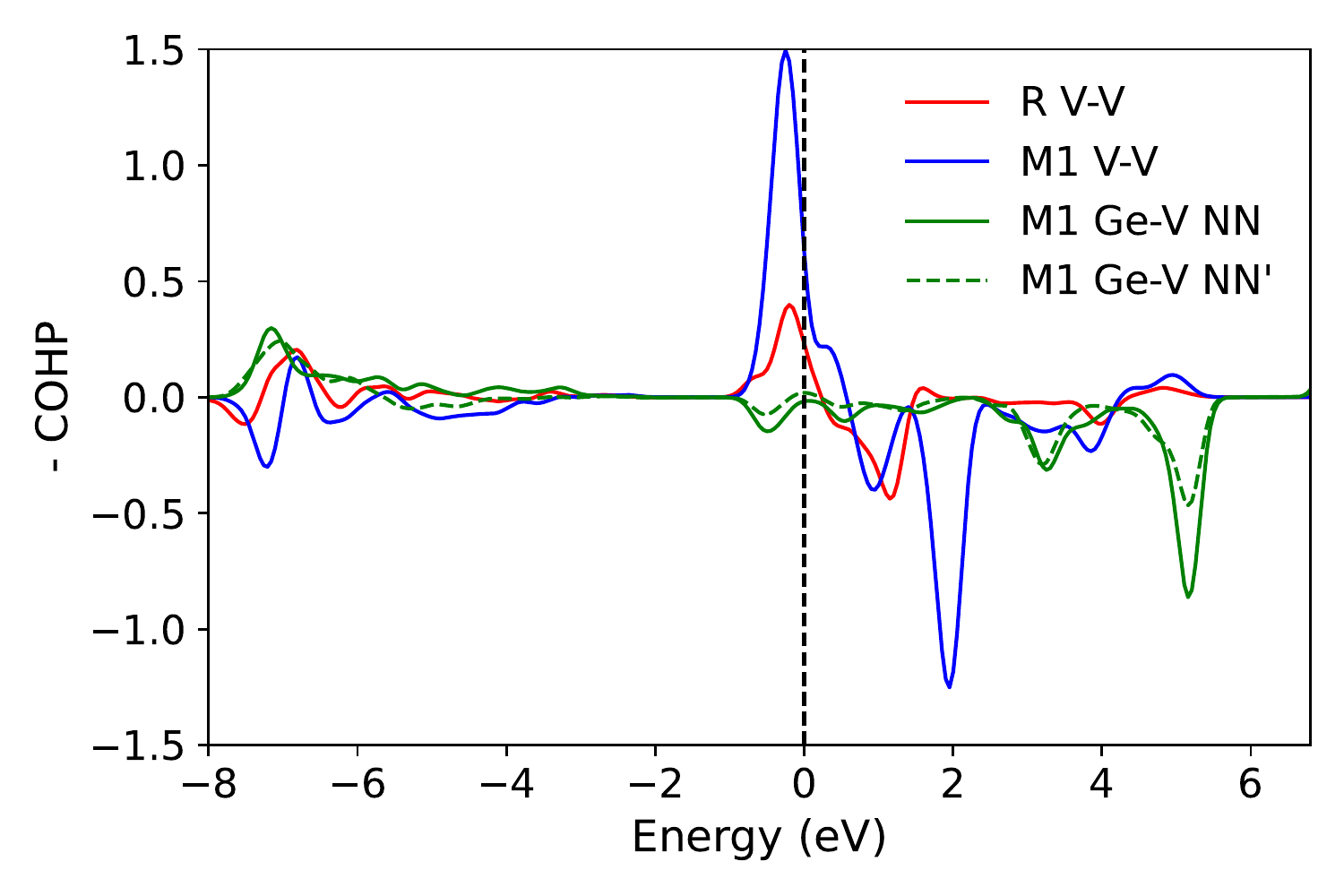}
    \caption{V-V and Ge-V COHPs in the R and M1 structures. The blue (red) line shows the undoped M1 (R) V-V bond behavior. Both peak near the Fermi level (black dashed line), with predominantly bonding (positive) character below it and antibonding (negative) character above it. This behavior is strongly enhanced in the M1 phase. In green, the Ge-V COHP in the Ge-doped M1 $2 \times 2 \times 2$ supercell is shown for both the formerly dimerized nearest neighbor (solid line), and the nearest neighbor along the chain in the opposite direction (dashed line); both COHPs show similar weak bonding.}
    \label{fig:cohp}
\end{figure}

Next, we examine the bonding behavior of the dopant in the supercell by calculating its COHP for the fully optimized structures (Fig.~\ref{fig:cohp}). The COHP of the undoped M1 phase displays a strong bonding-antibonding feature related to the $a_{1g}$ orbital (Fig.~\ref{fig:cohp}, blue), due to the dimerization of two V atoms. Such a strong feature is missing in the R phase, where the bonding behavior is much weaker (Fig.~\ref{fig:cohp}, red line ). We also show COHPs for the Ge-doped M1 system in Fig.~\ref{fig:cohp} (the R system behaves similarly although the relative changes are less strong). We see that the Ge-V bonding is significantly weaker than the previous V-V bonding. This is true for both the formerly dimerized nearest neighbor (NN), and also the nearest neighbor along the chain in the opposite direction (NN'), corresponding to a different dimer (dashed and solid green lines in Fig.~\ref{fig:cohp}). Both atoms bond weakly with the Ge atom, consistent with the absence of Ge-V dimerization and confirming the electronic inertness of the Ge dopant as seen through the band structure.

\subsubsection{Structure}

\begin{figure}[!t]
    \centering
    \includegraphics[width=0.95\linewidth]{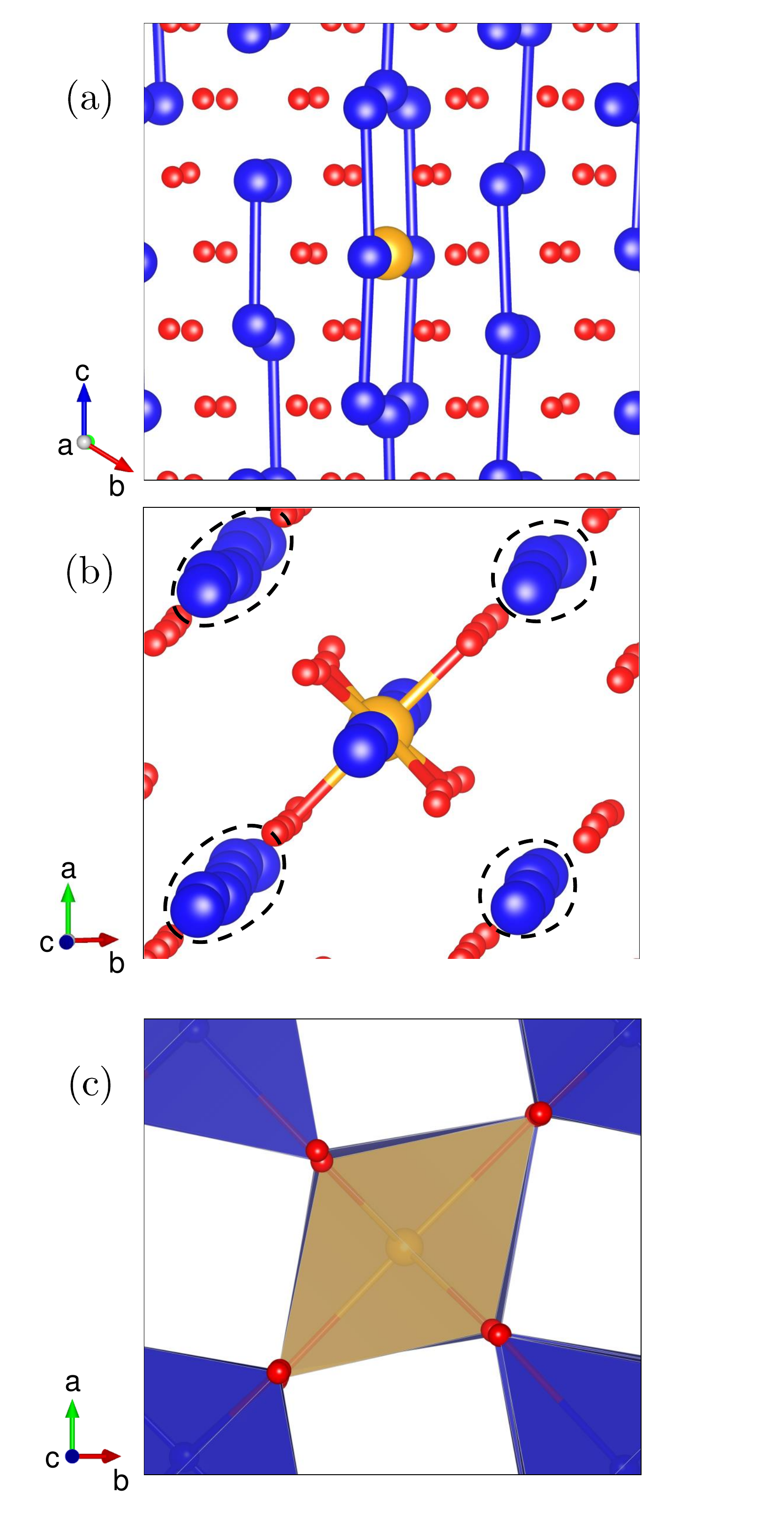}
    \caption{Details of the single-Ge-doped structurally relaxed $2 \times 2 \times 2$ R VO$_2$ supercell. (a) View along the $\vec{a}$ lattice parameter direction. V-V bonds are indicated for V-V distances less than 2.7~\r{A}. (b) View along the $\vec{c}$ direction. Buckled chains described in the text are highlighted with dashed ellipses. V, O, and Ge atoms shown in blue, red, and orange, respectively. Ge-O bonds are indicated for Ge-O distance less than 2.0~\r{A}. (c)~View along the $\vec{c}$ direction, now with oxygen polyhedra shown. Ge (V) centered polyhedra are shown in orange (blue).}
    \label{fig:structures_detail}
\end{figure}

Finally, we investigate the effect of substituting a single Ge atom into the $2 \times 2 \times 2$ VO$_2$ supercell on the structure of the two phases. After a structural relaxation following the introduction of the dopant Ge atom, we qualitatively observe very little difference between the doped and undoped M1 phases. The R phase, however, undergoes a significant distortion from its original high symmetry; see Fig.~\ref{fig:structures_detail} for details of the structurally relaxed supercell. The addition of a Ge atom has strong structural effects, with the V atoms next to the Ge dopant displaced away from it along the chain direction (Fig.~\ref{fig:structures_detail}~(a)). The GeO$_6$ polyhedron distorts the lattice and causes buckling of the neighboring chains (Fig.~\ref{fig:structures_detail}~(b)). Compared to the VO$_6$ polyhedron, the equatorial oxygens in GeO$_6$ are closer to the dopant, although the apical oxygens' Ge-O distances are unchanged, and the Ge remains at the center for the octahedron (Fig.~\ref{fig:structures_detail}~(c)).
 
The resulting structure forms zig-zagged chains (note the ellipses in Fig.~\ref{fig:structures_detail}~(b)), not unlike the pristine M1 phase, with some V atoms relaxing into positions closer to each other than before. The structural dimers that form, indicated as bonds for V-V distances less than 2.7~\r{A} in Fig.~\ref{fig:structures_detail}, with the smallest distances reaching 2.6~\r{A}, are similar to those found in the M1 phase. For a more quantitative discussion using histograms of nearest-neighbor distances, also featuring typical R and M1 V-V distances, see Fig.~\ref{fig:histograms}. We, however, also note that due to the periodicity effects coming from the construction of the $2 \times 2 \times 2$ supercell, the distortions along the chain with the added Ge atom also lead to the formation of ``trimers" (as seen in Fig.~\ref{fig:structures_detail} (a)) throughout the cell, since there are exactly three V atoms between subsequent Ge atoms.  

\subsection{VO$_2$ cells containing multiple Ge atoms}

Due to the potential constraints related to the limited size of the periodic $2 \times 2 \times 2$ supercell, resulting in the formation of the ``structural trimers'' seen in Fig.~\ref{fig:structures_detail}(a), we now consider a larger $2 \times 2 \times 3$ supercell with two Ge atoms. This treatment allows us to vary the relative positions of the dopant atoms and to check for spurious effects due to the periodic boundary conditions, especially along the Ge$_{0j}$ chain. In particular, with two Ge atoms, we are able to study multiple configurations with both even and odd numbers of V atoms between subsequent Ge atoms.

\begin{figure}
    \centering
    \includegraphics[width=1\linewidth]{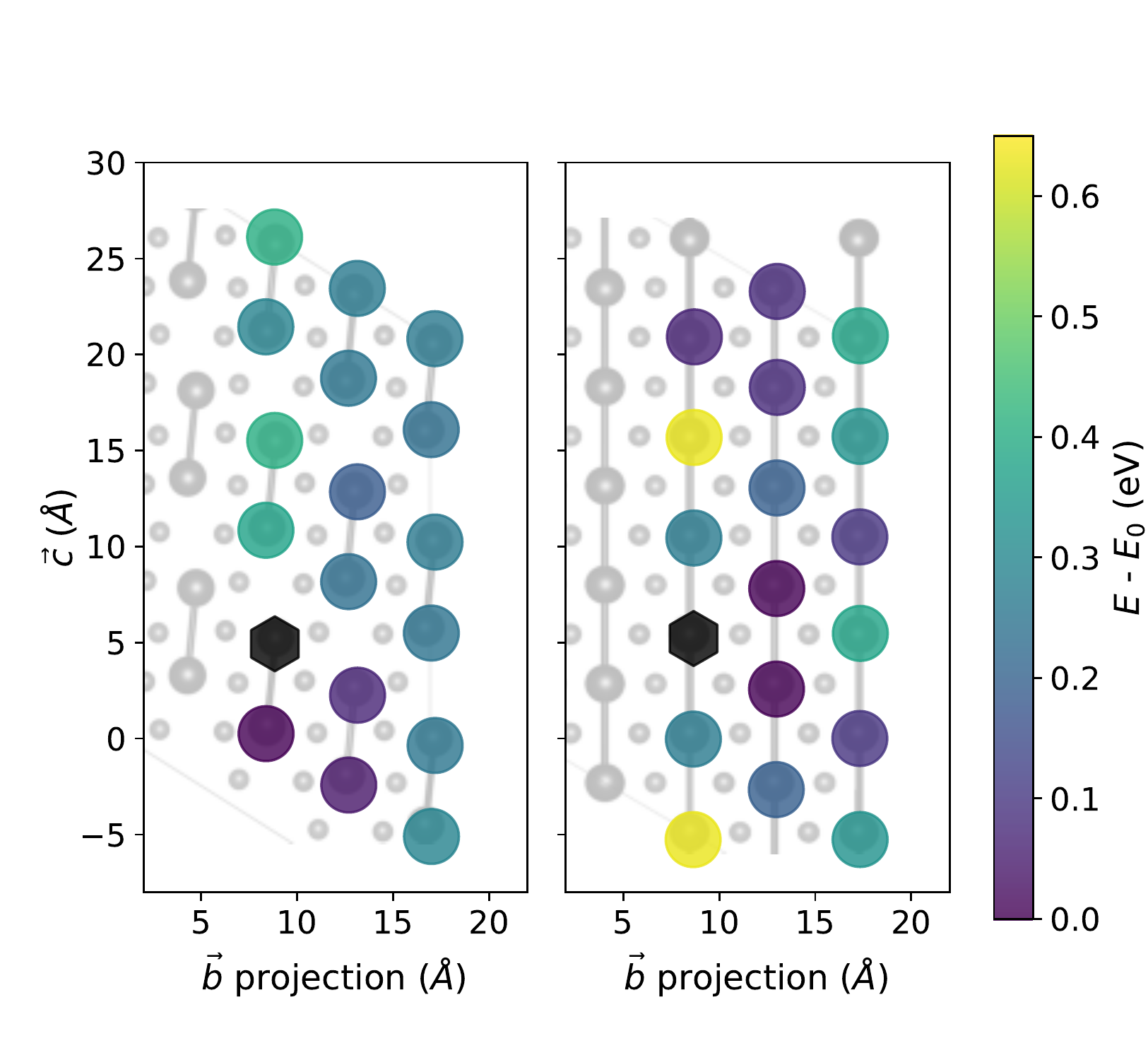}
    \caption{Energies of supercells of the M1 (left panel) and R (right panel) VO$_2$ phases when doped at the indicated symmetry inequivalent positions, as viewed down the $\vec{a}$ axis. Black hexagons correspond to the positions of one of the Ge atoms in the supercell. Color mapping indicates the total energy of the relaxed supercell with the second Ge atom placed at the indicated position, relative to the lowest energy ($E_0$) M1 (left panel) and R (right panel) configurations.}
    \label{fig:energetics}
\end{figure}

\subsubsection{Energetics}

We first consider the energetics of the different configurations of the two Ge atoms in the M1 phase after the structural relaxation (Fig.~\ref{fig:energetics}, left panel, overlaid on top of the unit cell structure). We begin by comparing the different relative energies of the relaxed structures. The lowest energy ($E_0$) occurs when the Ge atoms cluster, with a V-V dimer replaced by two Ge atoms (Ge$_{0-1}$). The second-lowest energy configurations have the second Ge atom in the nearest-neighbor chain (Ge$_{10}$, Ge$_{1-1}$), and are within 0.1~eV of $E_0$. In addition, the Ge$_{03}$ position, allowing even numbers of V atoms in between Ge dopants, is lower in energy than its neighboring configurations with 0.29~eV above $E_0$ compared to 0.40~eV for the surrounding positions within the chain. In this case, although dimerization is disfavored due to surrounding dimers and the zig-zag distortion orientation, the previously non-dimerized V atoms between the Ge atoms are also pushed closer together. The remaining Ge positions have similar energies at around 0.4~eV above $E_0$ (note the similar colors in Fig.~\ref{fig:energetics}, left panel).

\begin{figure*}
    \centering
    \begin{minipage}{0.48\textwidth}
        \centering
        \includegraphics[width=1\textwidth]{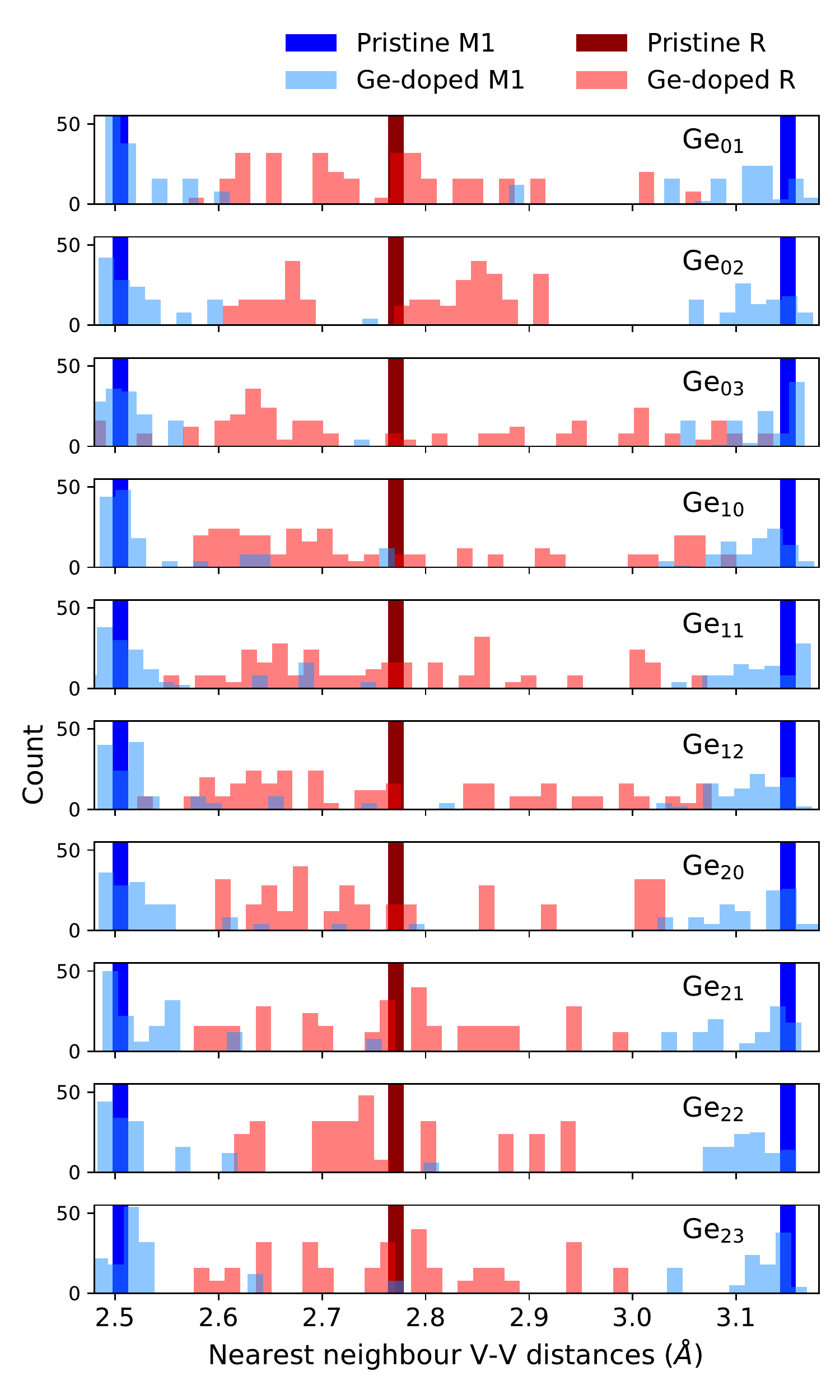}
        \caption{Nearest neighbor V-V distances for the various configurations upon doping in both R (red bars) and M1 (blue bars) phases. The original undoped nearest neighbor V-V distances are shown in dark colors, and the distances in the indicated structure after doping and relaxation are shown in lighter colors. The M1 distances only broaden, while the R phase distorts strongly.}
        \label{fig:histograms}
    \end{minipage}\hfill
    \begin{minipage}{0.48\textwidth}
        \centering
        \includegraphics[width=1\textwidth]{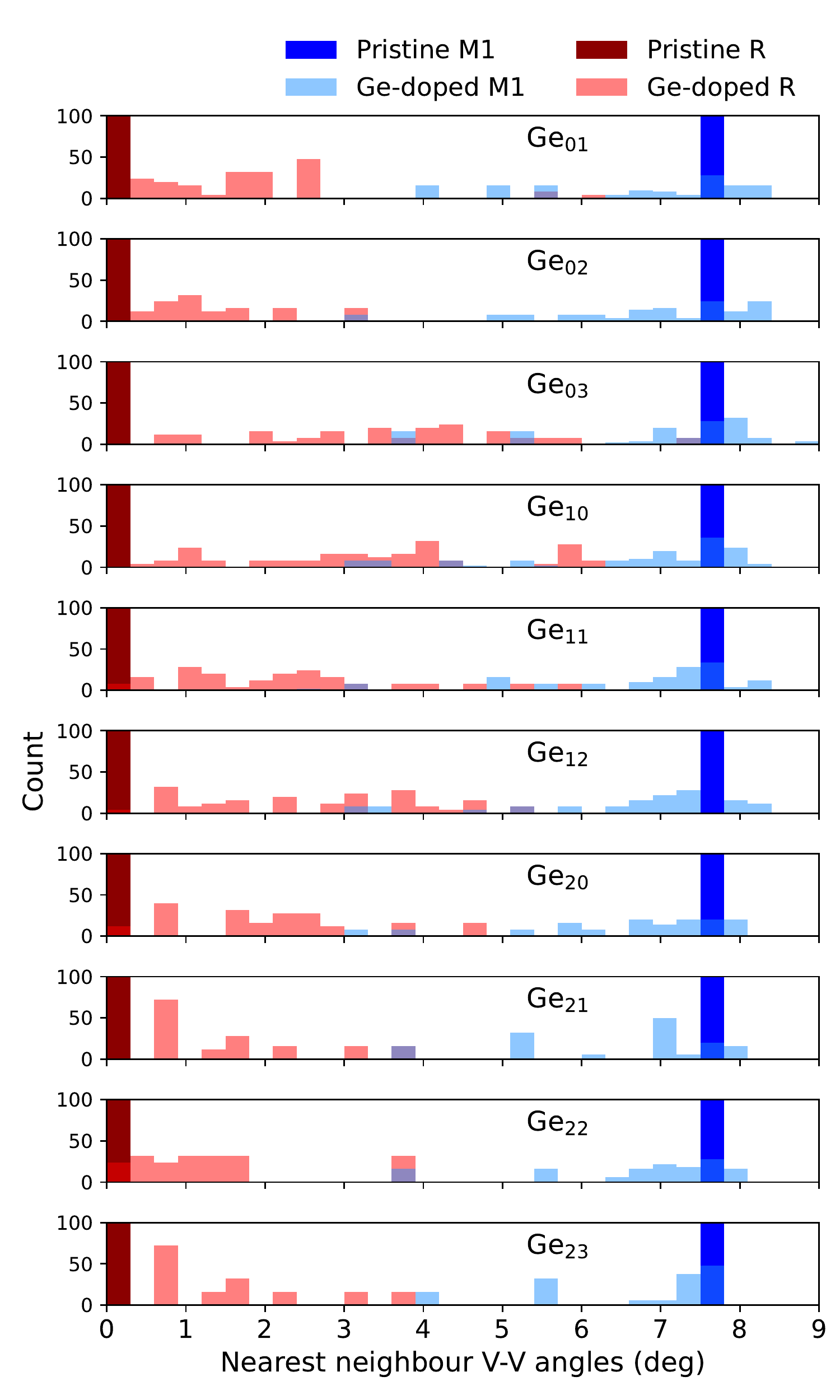}
        \caption{Nearest neighbor V-V angles for the various configurations upon doping in both R (red) and M1 (blue) phases. The original undoped nearest neighbor V-V angles are shown in dark colors, and the angles in the indicated structure after doping and relaxation are shown in lighter colors. Both the M1 and R angle distributions shift towards each other.}
        \label{fig:histograms_angles}
    \end{minipage}
\end{figure*}

Considering the R phase (Fig.~\ref{fig:energetics}, right panel), the calculated energies for the various Ge positions also show that clustering of Ge atoms in the nearest-neighbor chain (especially at Ge$_{11}$, and Ge$_{10}$ --- both the $E_0$ configurations) is the most favorable. However, in the R phase, the tendency for Ge atoms to disfavor an odd number of V atoms between Ge atoms in a chain is surprisingly more strongly pronounced than in the M1 phase. Leaving an unpaired lone V atom in between two Ge atoms leads to a high energy configuration even though there was originally no dimerization present in the R phase --- the Ge$_{03}$ corresponds to almost the same low energy as Ge$_{11}$, whereas Ge$_{02}$ is a highly unfavorable position for the Ge atom, 0.63~eV higher than $E_0$ (note the bright yellow color in Fig.~\ref{fig:energetics}, right panel). This trend is consistent with the results for the M1 phase, where, again, the Ge atoms allow V atoms to dimerize only if an even number of them are present in a given chain between two Ge dopants. The trends in the other chains are not as prominent, but all low-energy configurations can be seen to support some form of zig-zag formation in the structurally relaxed cell as in the single-Ge-substituted case.

\subsubsection{Structure}

To quantify how Ge substitution distorts the structure, we consider the distances and angles between nearest neighbor V atoms along the $c$ direction.
Figure~\ref{fig:histograms} shows histograms of the distances and Fig.~\ref{fig:histograms_angles} shows histograms of the angles (relative to the $c$ axis) of the nearest V-V pairs for all considered configurations of the Ge atoms in the doped cases, together with the corresponding values in pristine VO$_2$. In the unperturbed M1 structure, two nearest V-V distances are present, at 2.5~\r{A} and 3.15~\r{A}, forming an angle of 7.65$^{\circ}$ with the $c$ direction indicating the perfectly dimerized structure. The unperturbed R phase has a single peak at a distance of 2.76~\r{A}, and a single peak at 0$^{\circ}$ of the angle, corresponding to the equidistant V atoms along the $c$ direction.

Ge doping does not strongly affect the structure of the M1 phase for any of the considered configurations of the Ge atoms. From the histograms of nearest-atoms distances (Fig.~\ref{fig:histograms}, blue bars) in the Ge-doped supercell, we find only minimal perturbation from the undoped VO$_2$ M1 phase for all configurations. The nearest atom distances retain a double-peaked distribution, corresponding to the typical short and long DFT dimerization distances, respectively, indicating that dominant dimerization is still present in the structure. Broadening around the equilibrium distances is present but only rarely do some distances become similar to those in the R~phase. The same behavior is observed when nearest-neighbor angles are considered (Fig.~\ref{fig:histograms_angles}, blue bars). Under doping, the angle of the pristine crystal structure is essentially retained with the value of the angle becoming broadened.

In contrast to the M1 phase, in the R phase, Ge doping causes a severe structural distortion. The previously single-valued nearest-atom distance in the pristine R VO$_2$ crystal either splits into a two peaked distribution (e.g.~Ge$_{02}$), or demonstrably widens for all the different configurations of two Ge atoms, essentially spanning the whole range of distances between the dimerized and non-dimerized distances of the M1 structure (Fig.~\ref{fig:histograms}, red bars). There are large shifts of atomic positions from those of the undoped structure, with atoms in all chains including those far from the impurity atoms, affected by the introduction of Ge dopants. This shift is observed also in the nearest-neighbor angles (Fig.~\ref{fig:histograms_angles}, red bars). The angles between nearest neighbors become non-zero and show a wide distribution of angles for all configurations considered, even reaching values of up to 7$^\circ$, i.e.,  similar to that in the pristine M1 phase. This leads to the same effects seen above in the $2 \times 2 \times 2$ supercell with one Ge atom (see Fig.~\ref{fig:structures_detail} (b)), where we observed buckling of the V-V chains and their consequent structural dimerization. Due to the presence of two Ge atoms, in some configurations (e.g.~Ge$_{03}$) this buckling closely resembles that of the M1 phase, with dimerization effects and a shift in the angle of the respective dimers away from zero. 

\begin{figure}
    \centering
    \includegraphics[width=1\linewidth]{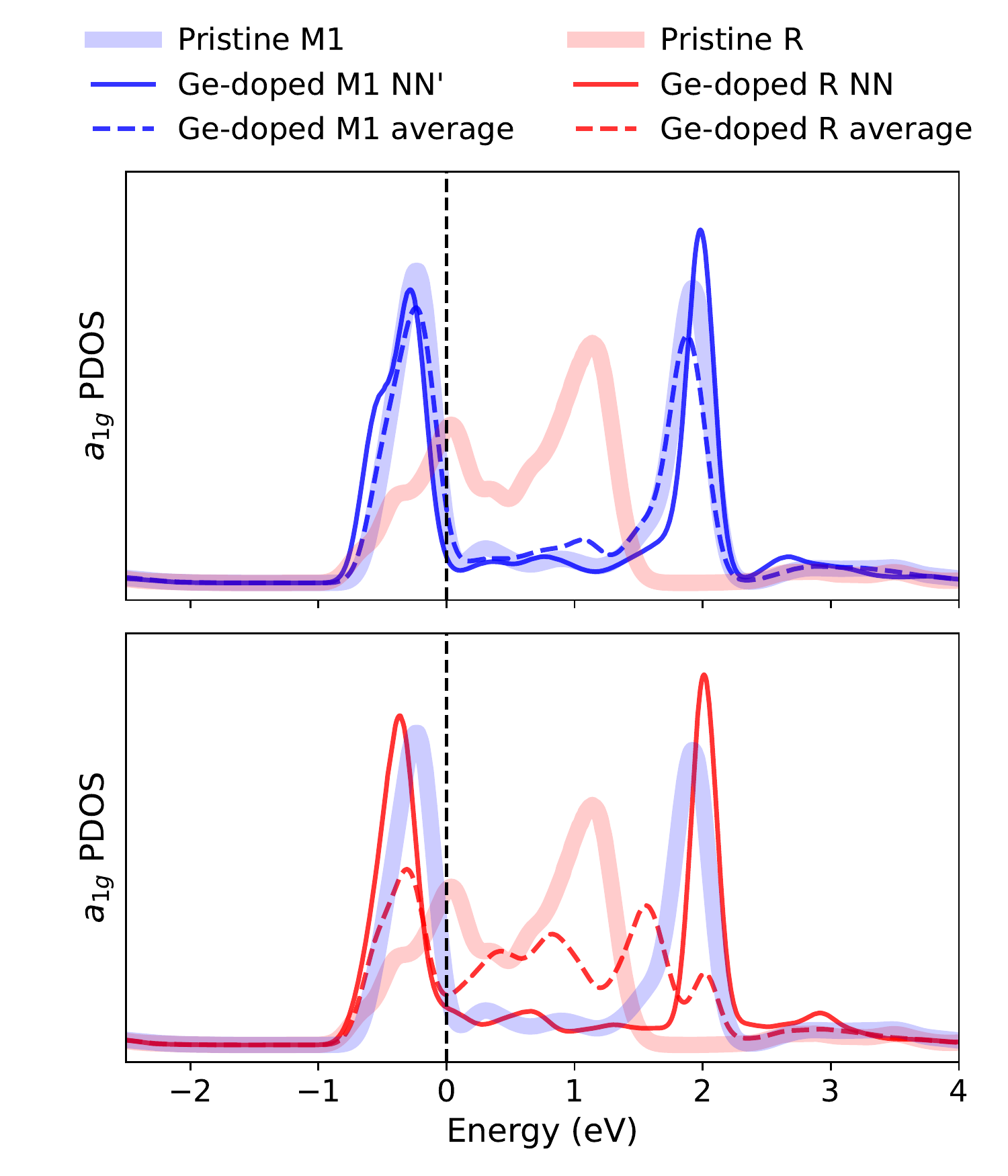}
    \caption{The PDOS onto the $a_{1g}$ orbital in the M1 (top panel) and R (bottom panel) Ge-doped supercells for the Ge$_{03}$ configuration. The shaded thick blue (red) lines respresent the pristine M1 (R) phases are shown. The solid line corresponds to the nearest neighbor to the Ge dopant atom, and the dashed line shows the average of all V atoms. The black dashed line corresponds to the Fermi level, set to 0~eV. The R phase PDOS is severely perturbed from the original state, while the M1 phase shows no significant distortion.}
    \label{fig:pdos}
    \vspace{1cm}
\end{figure}

\subsubsection{Electronics and chemical bonding}

Accompanying the structural distortions described above, the Ge-doped supercells also undergo electronic changes. The key change occurs in the $a_{1g}$ orbital projected density of states (PDOS), presented in Fig.~\ref{fig:pdos}.

In the M1 phase, we observe that the local DOS of the V atoms that remain dimerized is almost unchanged on Ge doping, as shown in the $a_{1g}$ PDOS for the Ge$_{03}$ configuration (Fig.~\ref{fig:pdos}, top panel). The average PDOS in doped systems still exhibits the typical bonding-antibonding feature of pristine M1 VO$_2$, as previously discussed in the context of the COHP (Fig.~\ref{fig:cohp}), and shown in Fig.~\ref{fig:pdos} as the thick blue line. The NN' (as defined in Sec.~\ref{sec:chemistry}) PDOS also keeps the shape of the undoped material with only minor deviations (note that we consider the nearest neighbor in the other direction in the chain, because the nearest-neighbor PDOS displays a rotated local coordinate system). This result is consistent with the observations of the structural distortions.

In contrast, in the rutile phase, the electronic structure changes significantly compared with the undoped case (Fig.~\ref{fig:pdos}, bottom panel, thick red line) on Ge incorporation. Both of the V nearest neighbors to the Ge dopant become electronically extremely similar to V atoms in the M1 structure, displaying a double-peaked bonding-antibonding PDOS for the Ge$_{03}$ configuration (see solid line in Fig.~\ref{fig:pdos}). Additionally, large changes also occur throughout the bulk of the supercell, and also the average $a_{1g}$ PDOS changes significantly from that of the undoped R phase (dashed line in Fig.~\ref{fig:pdos}). The overall average PDOS broadens, and its value at $E_F$ is reduced.


\section{Summary and Outlook}

In this work, we present a DFT study of the structural and electronic properties of Ge-doped VO$_2$ in its M1 and R structures.

We show that the electronic states of Ge atoms are not present near the Fermi level for the doped VO$_2$ phases considered. We further show that the Ge-V bonding is weak in both cases. Energetically, we observe that in both phases, the lowest energy arrangements of two Ge atoms correspond to their clustering close to each other. In both phases, the lowest energy arrangements correspond to configurations that lead only to small disruptions of the V-V pairs (since, except for Ge$_{0-1}$, all configurations do of course disrupt the V-V pairs to some extent).

However, we observe drastically different structural relaxations for the R and M1 phases. We find that the M1 phase is largely unperturbed both structurally and electronically by Ge doping. In contrast, the addition of Ge dopants to the R phase seems to push the structure towards the M1 dimerized phase. The structural distortion caused by the Ge atoms promotes structural dimer formation of the V atoms, as the buckling of the chains causes a tilt and alters the V-V distances and angles of these dimerized distances. Coupled with this are electronic changes in the R phase, in which the dimers cause an enhanced bonding-antibonding splitting in the DOS. 

We conclude that the R phase is more prone to structural perturbations caused by the Ge atoms than the M1 phase, hinting at an intrinsic instability of this phase. Our results also suggest that it is the strength of the dimers that allows the robustness of the M1 phase with respect to Ge incorporation. We further note that the large structural perturbations induced in the R phase on Ge doping give it a tendency towards insulating behavior, reducing the DOS at the Fermi level, again in the direction of the M1 phase. The strong distortions in the VO$_2$ rutile phase are remarkable, given that the ground state structure of GeO$_2$ is also the rutile structure. The behavior is reminiscent of the case of Ti$_4$O$_7$~\cite{Eyert/Schwingenschloegl/Eckern:2004, Leonov_et_al:2006}, in which the TiO$_2$ rutile layers, which are interleaved with layers containing $d^1$ Ti$^{3+}$ ions, are strongly distorted.

Therefore, our results hint toward a possible explanation of the recent experimental observation of $T_c$ increase under Ge doping. First, under doping we observe the R phase to be more structurally perturbed than M1, with the doping resulting in structural distortions that resemble those present in the pristine M1 phase. Second, we observe that the M1 phase is largely robust towards Ge incorporation, which only leads to minor structural distortions. This indicates that the low-temperature M1 phase is favored by Ge doping leading in turn to an increase in the transition temperature. However, to more persuasively arrive at these conclusions and to verify the hints presented here, further research with more advanced methods is required.

\section*{Acknowledgments}
We are thankful to Adrian Ionescu and Daesung Park for useful discussions and comments. Calculations were performed on the ETH Z\"{u}rich Euler cluster. This work was supported by ETH Z\"{u}rich. All input files available in Ref.~\cite{[{All input files available at }][{.}]supp}.

\bibliography{ge-doping}

\end{document}